\documentclass[floatfix,twocolumn,aps,prl,showpacs,amsmath,amssymb]{revtex4-1}
\usepackage[latin1]{inputenc}
\usepackage[dvips]{graphicx}

\usepackage{dcolumn}
\usepackage{bm}
\usepackage{dsfont}
\usepackage{color}
\usepackage{url}
\usepackage[colorlinks=true ,citecolor=blue]{hyperref}
\usepackage{amsmath,amssymb,esint}
 \usepackage[table]{xcolor}

\definecolor{Nathanblue}{rgb}{0.,0.24,0.51}

\newcommand{\be}{\begin{equation}}
	\newcommand{\ee}{\end{equation}}
\newcommand{\bq}{\begin{eqnarray}}
	\newcommand{\eq}{\end{eqnarray}}

\begin{document}

\title{Fermion-fermion duality in 3+1 dimensions}

\author{Giandomenico Palumbo}
\affiliation{Center for Nonlinear Phenomena and Complex Systems,
	Universit\'e Libre de Bruxelles, CP 231, Campus Plaine, B-1050 Brussels, Belgium}

\date{\today}

\begin{abstract}

Dualities play a central role in both quantum field theories and condensed matter systems.
Recently, a web of dualities has been discovered in 2+1 dimensions. Here, we propose in particular a generalization of the Son's fermion-fermion duality to 3+1 dimensions. We show that the action of charged Dirac fermions coupled to an external electromagnetic field is dual to an action of neutral fermions minimally coupled to an emergent vector gauge field. This dual action contains also a further tensor (Kalb-Ramond) gauge field coupled to the emergent and electromagnetic vector potentials. We firstly demonstrate the duality in the massive case. We then show the duality in the case of massless fermions starting from a lattice model and employing the slave-rotor approach already used in the 2+1-dimensional duality [Burkov, Phys. Rev. B \textbf{99}, 035124 (2019)]. We finally apply this result to 3D Dirac semimetals in the low-energy regime.
Besides the implications in topological phases of matter, our results shed light on the possible existence of a novel web of dualities in 3+1-dimensional (non-supersymmetric) quantum field theories.

\end{abstract}

\maketitle

\section{Introduction}

Dualities are powerful and non-perturbative relations between apparently different theories. They have a long history in both high-energy and condensed matter physics \cite{Polchinski,Savit}.
Dual models allow us to understand many aspects that are not easily accessible by analyzing directly the original starting models. 
Among them, in high-energy physics we can remind the AdS/CFT correspondence \cite{Maldacena}, S-duality \cite{Witten}, T-duality \cite{Strominger} and the more recent Chern-Simons-matter dualities \cite{Aharony,Giombi}.\\
In condensed matter physics, one of the most well studied duality is the particle-vortex duality that relates two different bosonic systems in 2+1 dimensions \cite{Peskin,Halperin}.
There have been recently a profound understanding of 2+1-D dualities in topological systems thanks to the seminal work by Son \cite{Son}, who proposed a novel fermion-fermion duality in the fractional quantum Hall effect. Here, a charged Dirac fermion coupled to an external electromagnetic field is shown to be dual to a composite neutral Dirac fermion coupled to an emergent vector gauge field, which is coupled to the electromagnetic potential through a topological BF term.
A more rigorous derivation of this correspondence on the lattice has been carried out in some papers \cite{Alicea,Burkov}.
Besides, the quantum Hall effect, this duality has been employed to study the surface of three-dimensional topological insulators \cite{Senthil,Vishwanath}. A web of dualities has been then unveiled with many non-trivial relations between free and interacting theories with bosons and fermions \cite{Seiberg,Tong,Kachru,Tong2,Nastase,Benini,Raghu,Koma,Raghu2,Xu,Benvenuti,Agarwal,Argurio}. Some of these results are deeply connected to higher-dimensional bosonization \cite{Tong, Jensen, Nastase2, Tong3}.
Thus, these dualites do not have only important impact in condensed matter systems, but have also relevant implications in high-energy physics.
Because both gauge fields and effective relativistic fermions are ubiquitous in topological phases of matter in any dimension \cite{Fradkin}, it is then natural to wonder if a web of dualities can exist in higher dimensions. Very recent results in 3+1-D bosonization and boson-fermion duality suggest that this is the case \cite{Ryu,Palumbo,Karch,Senthil2,Kapustin,Nishida}.

In this work, we provide a novel fermion-fermion duality in 3+1 dimensions by generalizing the Son's duality \cite{Son}. We will show that the action of free charged Dirac fermions coupled to an external electromagnetic field is dual to an action containing composite neutral fermions coupled to an emergent vector gauge field together and a tensor (Kalb-Ramond) gauge field \cite{Kalb} coupled to both the emergent and electromagnetic fields through a topological BF term. Differently from the lower-dimensional fermion-fermion duality, in our case the dual action contains two independent gauge fields. This can be naturally explained from the fact that the topological BF coupling in 3+1 dimensions \cite{Birmingham}, naturally involves a vector and tensor gauge fields. We remind that the B-field has many important applications in topological quantum field theories \cite{Birmingham,Horowitz}, string theory \cite{Kalb, Szabo,Banks} and condensed matter physics \cite{Hansson,Sodano,Moore,Maciejko,Gu,Tiwari,Cappelli,Zaanen,Wang}. Moreover, its momentum-space version has been recently introduced in band theory to define generalized Berry connections and tensor monopoles \cite{Palumbo2,Palumbo3}. 
The Kalb-Ramond field appears also in a 3+1-D version of the particle-vortex duality \cite{Franz} and bosonization \cite{Ryu,Palumbo,Kapustin,Botta,Quevedo,Marino,Fosco}.
Finally, we will show a natural application of our duality to 3D Dirac semimetals \cite{Rappe,Armitage}.
 
This work sheds light on the existence of novel and unsuspected dualities between apparently different fermion models in 3+1 dimensions with very relevant implications in both topological phases and high-energy physics.

\section{Duality in the massive case}

We start presenting our fermion-fermion duality in 3+1 dimensions, between free Dirac fermions $\psi$ coupled to an electromagnetic field $A_{\mu}$
\begin{eqnarray}\label{SA}
S_{1}[\bar{\psi},\psi, A_{\mu}]=\int_{M}  d^{4}x\, \bar{\psi}\, \Gamma^{\mu}(\partial_{\mu}+i A_{\mu})\psi,
\end{eqnarray}
with $\mu=\{x,y,z,t\}$, $M$ is the 3+1-D flat manifold, $\bar{\psi}=\psi^{\dagger}\Gamma^{0}$, $\Gamma^{\mu}$ are $4\times 4$ Dirac matrices,
and composite neutral Dirac fermions $f$ coupled to an emergent Abelian gauge field $a_{\mu}$
\begin{align}\label{SB}
S_{2}[\bar{f},f, A_{\mu}, a_{\mu},B_{\mu\nu}]=\int_{M}  d^{4}x\, \left[\bar{f}\, \Gamma^{\mu}(\partial_{\mu}+ia_{\mu})f \right.    \nonumber \\
\left. -(1/8\pi)\epsilon^{\mu\nu\lambda\delta} B_{\mu\nu}(F_{\lambda\delta}(A)+F_{\lambda\delta}(a))\right. \nonumber \\  \left.+(1/32\pi^{2}\chi)\mathcal{H}_{\mu\nu\lambda}\mathcal{H}^{\mu\nu\lambda}\right],
\end{align}
where $F_{\lambda\delta}(A)=\partial_{\lambda}A_{\delta}-\partial_{\delta}A_{\lambda}$, $F_{\lambda\delta}(a)=\partial_{\lambda}a_{\delta}-\partial_{\delta}a_{\lambda}$, $\mathcal{H}_{\mu\nu\lambda}=\partial_{\mu}B_{\nu\lambda}+\partial_{\nu}B_{\lambda\mu}+\partial_{\lambda}B_{\mu\nu}$ and $\chi$ is a real parameter. Here, $B_{\mu\nu}$ is an emergent anti-symmetric tensor gauge field coupled to both $a_{\mu}$ and $A_{\mu}$.
This tensor field, known as Kalb-Ramond field in the string-theory literature \cite{Kalb,Szabo} is Abelian and transforms as follows: $B_{\mu\nu}\rightarrow B_{\mu\nu}+\partial_{\mu}\zeta_{\nu} -\partial_{\nu}\zeta_{\mu}$, where $\zeta_{\mu}$ is a generic vector field that contains the gauge redundancy. Importantly, when $M$ is compact, the BF terms in Eq.(\ref{SB}) are gauge invariant \cite{Birmingham}.
Our proposal naturally generalizes the Son's fermion-fermion duality in 2+1 dimensions \cite{Son}. However, differently from the lower-dimensional case, in our dual action Eq.~(\ref{SB}) there appear two independent gauge fields. A couple of comments are necessary at this point. In the original fermion-fermion duality, one of the main ingredients is given by the presence of the 2+1-dimensional BF term (or mixed Chern-Simons term) that couples the emergent gauge field to the electromagnetic potential. In Eq.~(\ref{SB}) there appears the natural higher-dimensional version of this topological term. Importantly, the Kalb-Ramond field cannot couple directly to fermions due to a lacking of gauge invariance and for this reason there emerges a further U(1) vector gauge field $a_{\mu}$, which couples directly to the neutral fermion field. Importantly, the additional degrees of freedom in $S_{2}$ given by the kinematic term of the Kalb-Ramond field, are vortices, which in
3+1 dimensions behave like strings.
The demonstration of this duality for massless fermions on the lattice will be presented in the next section.

In the massive case instead the duality in the continuum can be proved in a straightforward way as we show now.
By adding a Dirac mass $m$, with $m<0$, in both actions in Eqs.~(\ref{SA}) and (\ref{SB}), we can integrate out the charged and neutral fermions, by obtaining the corresponding effective actions $S_{1}^{\rm eff}[A_{\mu}]$ and $S_{2}^{\rm eff}[A_{\mu}, a_{\mu},B_{\mu\nu}]$ respectively, in the low-energy (infrared) regime.
At leading order, the two effective actions contain topological theta terms \cite{Zhang,Hosur} for $A_{\mu}$ and $a_{\mu}$, respectively
\begin{eqnarray}\label{ST1}
S_{1}^{\rm eff}[A_{\mu}]=\frac{1}{32\pi}\int_{M}  d^{4}x\, \epsilon^{\mu\nu\lambda\delta} F_{\mu\nu}(A)F_{\lambda\delta}(A), \hspace{1.7cm}
\end{eqnarray}
\begin{align}\label{ST2}
S_{2}^{\rm eff}[A_{\mu}, a_{\mu},B_{\mu\nu}]=\int_{M}  d^{4}x\, \epsilon^{\mu\nu\lambda\delta}\left[ (1/32\pi) F_{\mu\nu}(a)F_{\lambda\delta}(a) \right. \nonumber \\
\left. -(1/8\pi) B_{\mu\nu}(F_{\lambda\delta}(A)+F_{\lambda\delta}(a))+(1/32\pi^{2}\chi)\mathcal{H}_{\mu\nu\lambda}\mathcal{H}^{\mu\nu\lambda}\right].
\end{align}
In the low-energy limit, the topological terms in $S_{2}^{\rm eff}$ are dominant. Thus, in the massive case we can neglect the kinematic term in Eq.~(\ref{ST2}) and the tensor gauge field behaves like a Lagrange multiplier field. By varying $S_{2}^{\rm eff}[A_{\mu}, a_{\mu},B_{\mu\nu}]$ with respect to $B_{\mu\nu}$ we obtain
\begin{align}
F_{\lambda\delta}(a)=-F_{\lambda\delta}(A).
\end{align}
By using this relation between the two vector fields in Eq.~(\ref{ST2}), we recover an action that depends only on $A_{\mu}$ and coincides with $S_{1}^{\rm eff}[A_{\mu}]$ in Eq.~(\ref{ST1}).
We remind that the topological theta term $S_{1}^{\rm eff}[A_{\mu}]$ describes the bulk of time-reversal-invariant 3D topological insulators in presence of an external electromagnetic potential \cite{Zhang}. On the other hand, the BF theory in Eq.~(\ref{ST2}) has been also employed to describe these topological phases \cite{Moore,Ryu}. Thus, our duality in the massive case naturally applies to these quantum systems, which are characterized on boundary by topologically protected massless Dirac fermions due to time-reversal symmetry. Here, we want to stress the fact that massive Dirac fermions play a central role in the duality because the theta terms in Eqs. (\ref{ST1}) and (\ref{ST2}) would not have been obtained in the case we had considered, for instance, non-relativistic fermions.

We show now that there exists a natural duality between the 2+1-D topological field theories defined on the gapped boundary of Eqs.~(\ref{ST1}) and (\ref{ST2}). We define $S_{1}^{\rm eff}[A]$ and $S_{2}^{\rm eff}[A_{\mu}, a_{\mu},B_{\lambda\delta}]$ on a slab geometry $M$ with open boundary $\partial M$. 
Massless Dirac modes on the boundary can be gapped by introducing, for instance, an external Zeeman field that breaks time-reversal symmetry. The gapped bulk of $S_{1}^{\rm eff}[A]$ is given by an Abelian Chern-Simons (CS) theory
\begin{eqnarray}\label{CS1}
S_{CS}[A_{\mu}]=\frac{1}{8\pi}\int_{\partial M}  d^{3}x\, \epsilon^{\mu\nu\lambda} A_{\mu}\partial_{\nu}A_{\lambda},
\end{eqnarray}
which has been derived by employing the Stokes theorem \cite{Cappelli}. This is the effective action for half-integer quantum Hall states. The topological boundary theory for $S_{2}^{\rm eff}[A_{\mu}, a_{\mu},B_{\mu\nu}]$ can be derived by following Ref.~\cite{Moore}. It is given by
\begin{align}\label{CS2}
S_{CS}[A_{\mu}, a_{\mu},b_{\mu}]=  \hspace{5.5cm}\nonumber \\
\frac{1}{8\pi}\int_{\partial M}  d^{3}x\, \epsilon^{\mu\nu\lambda} \left( a_{\mu}\partial_{\nu}a_{\lambda}-b_{\mu}\partial_{\nu}a_{\lambda}-b_{\mu}\partial_{\nu}A_{\lambda}\right),
\end{align}
where $b_{\mu}$ is a novel vector gauge field induced by the Kalb-Ramond field on the boundary. In fact, the BF terms are not gauge invariant on manifolds with boundary. This issue can be cured by introducing the CS terms in Eq.~(\ref{CS2}), such that the total system, bulk plus boundary, preserves the gauge invariance \cite{Moore}.
By integrating out first $a_{\mu}$ and then $b_{\mu}$, we obtain a Chern-Simons term that depends only on $A_{\mu}$ and coincides with Eq.~(\ref{CS1}).
Clearly, this correspondence holds because of the 3+1-D duality in the bulk.
In the next section, we will demonstrate the nontrivial duality in the massless case by starting with fermions on the lattice.

\section{Duality from a lattice model}
We start considering a three-dimensional tight-binding model on a cubic lattice for fermions coupled to an external electromagnetic field. The corresponding Hamiltonian is given by \cite{Zhang}
\begin{eqnarray}\label{Dirac}
H_{\psi}=\sum_{r,s}\left[\psi^{\dagger}_{r}\left(\frac{\xi\Gamma^{0}-i \Gamma^{s} }{2}\right)e^{i A_{r,r+\hat{s}}}\psi_{r+\hat{s}}\right]+{\rm h.c.} \\ \nonumber
+\sum_{r}\psi^{\dagger}_{r}\left[ \sin \phi\, \Gamma^{5}+(m+ \xi \cos \phi)\Gamma^{0}-i A_{r,0}\right]\psi_{r},
\end{eqnarray}
where $r=\{x,y,z\}$ is the site index, $\phi$ is an adiabatic parameter, $m$ and $\xi$ are real constant parameters, $A_{r,r+\hat{s}}$ is the electromagnetic field introduced through a Peierls substitution on the lattice link $(r, r+\hat{s})$ with $\hat{s}\equiv (\hat{x},\hat{y}, \hat{z})$, $\psi_{r}$ is a four-component spinor and $\Gamma^{\alpha}$ are the Dirac matrices, with $\Gamma^{5}=i \Gamma^{0}\Gamma^{1}\Gamma^{2}\Gamma^{3}$ the chiral matrix.
This Hamiltonian describes Dirac fermions on the lattice and the bulk properties of time-reversal invariant topological insulators. Originally, this model was introduced in lattice gauge theory to study relativistic fermions in the discrete, named also Wilson fermions \cite{Wilson,Goldman,Hauke}.
On the second line in Eq.~(\ref{Dirac}) we can recognize the usual Dirac mass proportional to $\Gamma^{0}$. In the continuum limit (low-energy regime) at $\Gamma$-point (with $\Gamma\equiv (0,0,0)$) for $\phi=0$ and $\xi\rightarrow0$, we obtain the standard Dirac Hamiltonian
\begin{align}\label{Dirac2}
H_{\psi}=\int d^{3}x\left[ \psi^{\dagger}\Gamma^{j}(\partial_{j}+i A_{j}) \psi +m \psi^{\dagger} \Gamma^{0}\psi-i A_{0}\psi^{\dagger}\psi\right],
\end{align}
with $j=\{x,y,z\}$.
We are interested here to derive the dual model of this free fermion system in the massless regime.
For this reason, we fix $m=0$ and take $\phi=0$ for simplicity in Eq.~(\ref{Dirac}).
This gapless system can describe 3D Dirac semimetals \cite{Armitage} and the implications of our results in these topological systems will be stressed in the next section.
By following Ref.~\cite{Burkov}, we employ the slave-rotor approach to derive the dual theory of Eq.~(\ref{Dirac}).
We decompose the fermion creator operator $\psi_{r}^{\dagger}$ as follows \cite{Georges}
\begin{eqnarray}\label{slave-rotor}
\psi_{r}^{\dagger}=e^{i \theta_{r}}f_{r}^{\dagger},
\end{eqnarray}
where $e^{i \theta_{r}}$ is the creator operator for spinless bosons and $f_{r}^{\dagger}$ is a fermion creator operator associated to neutral fermions.
The conjugate of $\theta_{r}$ is given by the number operator $n_{r}$, such that $[\theta_{r}, n_{r}]=i$.
Like in Ref.~\cite{Burkov}, we consider the situation in which the slave rotor field does not condense, namely $\langle e^{i \theta_{r}}\rangle\neq 0$. Importantly, the neutral fermion operator satisfies the following condition
\begin{eqnarray}\label{number}
f^{\dagger}_{r}f_{r}=n_{r}+1,
\end{eqnarray}
where there is no summation on the $r$ index. Here, $n_{r}=-1$ represents the state with no electrons on site $r$, $n_{r}=0$ is the two states with one electron
and finally $n_{r}=1$ is the state with two electrons at $r$.
By substituting Eq. (\ref{slave-rotor}) in the lattice Hamiltonian (\ref{Dirac}), we have
\begin{eqnarray}\label{DualDirac}
H_{f}=\hspace{6.0cm}   \\ \nonumber
\sum_{r,s}\left[f^{\dagger}_{r}\left((\xi\Gamma^{0}-i \Gamma^{s})/2\right)e^{i( A_{r,r+\hat{s}}+\Delta_{\hat{s}}\theta_{r})}f_{r+\hat{s}}\right. \\ \nonumber
+{\rm h.c.}+\left.\xi f^{\dagger}_{r}\Gamma^{0}f_{r} -i n_{r} A_{r,0}\right], \hspace{1.4cm}
\end{eqnarray}
where $\Delta_{\hat{s}}\theta_{r}=\theta_{r}-\theta_{r,r+\hat{s}}$.
The corresponding imaginary-time action ($\tau=i t$) is then given by
\begin{eqnarray}\label{action}
S=\int_{0}^{\beta}d\tau \sum_{r,s}\left[f_{r}^{\dagger}\partial_{\tau} f^{\dagger}_{r}-i n_{r}(\partial_{\tau}\theta_{r}+A_{r,0})+ \right. \hspace{0.4cm} \\ \left. \nonumber
 i \lambda_{r}(f_{r}^{\dagger}f_{r}-n_{r}-1)+\xi f^{\dagger}_{r}\Gamma^{0}f_{r}+ \right. \\ \left.\nonumber
f_{r}^{\dagger}\left((\xi\Gamma^{0}-i \Gamma^{s} )/2\right)e^{i( A_{r,r+\hat{s}}+\Delta_{\hat{s}}\theta_{r})}f_{r+\hat{s}} 
 +{\rm h.c.}\right],
\end{eqnarray}
where $\lambda_{r}$ is the Lagrange multiplier field included to impose Eq.~(\ref{number}) in the action.
To deal with the last term in Eq.~(\ref{action}), we introduce an Hubbard-Stratonovich field $h\equiv\chi e^{i a}$ defined on the lattice \cite{Lee}. By considering its magnitude $\chi$ constant, Eq.~(\ref{action}) can be rewritten as follows
\begin{eqnarray}\label{HS}
S= \hspace{7.0cm}   \\ \nonumber 
\int_{0}^{\beta}d\tau \sum_{r,s}\left[f_{r}^{\dagger}(\partial_{\tau}+i a_{r,0}) f^{\dagger}_{r}-i n_{r}(\partial_{\tau}\theta_{r}+A_{r,0}+a_{r,0})+ \right. \\ \left.\nonumber
\xi f^{\dagger}_{r}\Gamma^{0}f_{r}+
\chi f_{r}^{\dagger}\left((\xi\Gamma^{0}-i \Gamma^{s})/2\right)e^{i a_{r,r+\hat{s}}}f_{r+\hat{s}}+{\rm h.c.}  \right. \\ \left.\nonumber
+\chi \cos \left(\Delta_{\hat{s}}\theta_{r}+A_{r,r+\hat{s}}+a_{r,r+\hat{s}}\right) \right], \hspace{1.4cm}
\end{eqnarray}
where $a_{r,0}\equiv \lambda_{r}$.

After employing the Villain approximation for the last term in Eq.~(\ref{HS}) \cite{Burkov}, the action can be decomposed as follows
\begin{eqnarray}
S=S_{f}+S_{\theta},
\end{eqnarray}
where
\begin{eqnarray}\label{S1}
S_{f}=\int_{0}^{\beta}d\tau \sum_{r,s}\left[f_{r}^{\dagger}(\partial_{\tau}+i a_{r,0}) f^{\dagger}_{r} +\xi f^{\dagger}_{r}\Gamma^{0}f_{r}+  \hspace{1.3cm} \right. \\ \left.\nonumber
\chi f_{r}^{\dagger}\left((\xi\Gamma^{0}-i \Gamma^{s} )/2\right)e^{i a_{r,r+\hat{s}}}f_{r+\hat{s}}+{\rm h.c.}\right], \hspace{1.2cm}
\end{eqnarray}
\begin{eqnarray}\label{S2}
S_{\theta}=\int_{0}^{\beta}d\tau \sum_{r,s}\left[i J^{r,0}(\partial_{\tau}\theta_{r}+A_{r,0}+a_{r,0})+  \hspace{1.5cm} \right.  \\ \left.\nonumber
i J^{r,r+\hat{s}} \left(\Delta_{\hat{s}}\theta_{r}+A_{r,r+\hat{s}}+a_{r,r+\hat{s}}\right)+(1/2\chi)(J^{r,r+\hat{s}})^{2} \right],
\end{eqnarray}
where $J^{r,r+\hat{s}}$ is the bosonic current, while its temporal component $J^{r,0}$ is identified with $n_{r}$.
In the continuum limit (i.e. long-wavelength limit), Eq.~(\ref{S1}) becomes
\begin{eqnarray}
S_{f}=\int  d^{4}x\, \bar{f}\, \Gamma^{\mu}(\partial_{\mu}+ia_{\mu})f, 
\end{eqnarray}
where $\bar{f}=f^{\dagger}\Gamma^{0}$ and $\mu=\{x,y,z,t\}$. In this limit, $\Delta_{\hat{s}}$ becomes the standard spacial derivative and by integrating out $\theta$ in Eq. (\ref{S2}), we obtain
\begin{eqnarray}
\partial_{\mu}J^{\mu}=0.
\end{eqnarray}
A solution for this equation is given by
\begin{eqnarray}
J^{\mu}=\frac{1}{4 \pi}\epsilon^{\mu\nu\lambda\delta}\partial_{\nu}B_{\lambda\delta},
\end{eqnarray}
where $B_{\lambda\delta}$ is an antisymmetric tensor gauge field.
After integrating by parts, our dual action in the continuum acquires its final form
\begin{align}
S =\int_{M}  d^{4}x\, \left[\bar{f}\, \Gamma^{\mu}(\partial_{\mu}+ia_{\mu})f \right.    \nonumber \\
\left. -(1/8\pi)\epsilon^{\mu\nu\lambda\delta} B_{\mu\nu}(F_{\lambda\delta}(A)+F_{\lambda\delta}(a))\right. \nonumber \\  \left.+(1/32\pi^{2}\chi)\mathcal{H}_{\mu\nu\lambda}\mathcal{H}^{\mu\nu\lambda}\right],
\end{align}
which coincides with Eq.~(\ref{SB}). Importantly, the above kinematic term for the Kalb-Ramond field gives rise to (dynamical) string-like vortices, which are not gapped. This is also compatible with the absence of condensation of the slave rotor field as specified below Eq. (\ref{slave-rotor}).

Thus we have shown that a duality holds between Eqs.~(\ref{SA}) and (\ref{SB}). This represents the main result of this paper.

Finally, we can identify a vector field $\bar{a}_\mu$ associated to the vortex loop current as a linear combination of the Hubbard-Stratonovich and electromagnetic fields, namely $\bar{a}_\mu=a_\mu+A_\mu$.
A vortex loop current $J^{\mu\nu}$ can be introduced in the dual action as follows
\begin{align}
S\rightarrow S + J^{\mu\nu}B_{\mu\nu}.
\end{align}
By avoiding for simplicity the kinematic term for $B_{\mu\nu}$, which does not affect our conclusions, we can vary $S$ with respect to $B_{\mu\nu}$, by getting
\begin{align}
J^{\mu\nu}=\frac{1}{4 \pi}\epsilon^{\mu\nu\lambda\delta} \partial_{\lambda} \bar{a}_\delta,
\end{align}
where $J^{\mu\nu}$ is clearly conserved. This show that $\bar{a}_\mu$ is the physical field associated to the vortex loop current at ground state.

\section{3D Dirac semimetals}
 After the discovery of graphene \cite{Novoselov}, there have been many efforts to find out higher-dimensional quantum systems supporting massless Dirac quasiparticles \cite{Armitage}. Dirac semimetals are novel topological semimetals with many unique features \cite{Rappe}. These systems are both time-reversal and inversion invariant, supporting a Dirac point, which is protected by spacial group symmetries. 
 They are described in the low-energy regime, by a three-dimensional massless Dirac Hamiltonian in the bulk given by Eq.~(\ref{Dirac2}) with $m=0$. Moreover, they can be built by stacking two-dimensional
 quantum spin Hall insulators in the momentum space along, for instance, the $k_{z}$ direction.
  By applying an external electromagnetic field in this system, one can derive the topological response of the bulk due to the external probing, such as the chiral anomaly \cite{Xiong}.

In this context, our duality implies that the free-fermion model describing Dirac semimetals can be mapped to an interacting phase, in which there appear a composite Dirac field, an emergent static vector gauge field $a_{\mu}$ and a dynamical Kalb-Ramond field $B_{\mu\nu}$. This is analogous to the 2+1-D fermion-fermion duality on the boundary of 3D topological insulators, where point-like vortices emerge in the dual action \cite{Senthil,Vishwanath}. However, in our case, the tensor gauge field gives rise to string-like vortices in the bulk, which could carry fractional statistics in 3+1 dimensions \cite{Wang}.

 Interactions in Dirac semimetals have been intensively analyzed in several works \cite{Roy,Herbut,Gonzalez}.
 With strong repulsive interactions, the system becomes gapped due to the chiral symmetry breaking that gives rise to a dynamical mass generation. On the other hand, attractive interactions can give rise to superconducting ground states.
 The Dirac semimetal is robust against sufficiently weak electron-electron interactions because the 
 density of states vanishes at the touching point (i.e. Dirac point). 
 Thus, our massless dual action in Eq.~(\ref{SB}) holds in this weakly-interacting limit. In particular, from the variation of $S_{2}[\bar{f},f, A_{\mu}, a_{\mu},B_{\mu\nu}]$ with respect to $a_{\mu}$ we have that
 \begin{eqnarray}
 i\bar{f}\,\Gamma^{\mu}f=\frac{1}{4 \pi}\epsilon^{\mu\nu\lambda\delta}\partial_{\nu}B_{\lambda\delta},
 \end{eqnarray}
 where the conserved current $j^{\mu}=i\bar{f}\,\Gamma^{\mu}f$ describes the low-energy matter fluctuations and is expressed in terms of the B-field, which plays the role of hydrodynamic gauge field for string-like vortices \cite{Cappelli}.
 Finally, the duality shows that 3D Dirac semimetals are not only topological at the level of the momentum-space band structure but are topological even in their low-energy description in real space.
 
 \section{Conclusions}
In this paper, we have proposed a novel duality between two apparently different fermion models in 3+1 dimensions.
One model is defined by a charged Dirac field coupled to an external electromagnetic potential, while the second one is given in terms of a neutral fermion field coupled to an emergent Abelian vector gauge field. This second model contains also an emergent tensor gauge field coupled to both the emergent and external vector potentials. This fermion-fermion duality represents the natural generalization of the 2+1-dimensional duality proposed by Son. The demonstration of the correspondence in the massive case is straightforward while the non trivial massless case requires a more sophisticated approach on the lattice through the slave-rotor formalism \cite{Burkov}. 
The proof of the duality for massless Dirac fermions by using uniquely quantum-field-theory techniques in the continuum will be presented in future work.

Our results do not have only relevant applications in topological phases of matter, but shed new light on dualities in higher dimensions making a glimpse of a novel web of dualities. In this new scenario,  the fermion-fermion duality together with the recently discovered boson-fermion duality \cite{Nishida} would give rise, for instance, to a novel higher-dimensional boson-boson duality, with very relevant implications in both condensed matter and high-energy physics.

\vspace{0.5cm}

 {\bf Acknowledgments:} We thank Riccardo Argurio and Pierluigi Niro for discussions and relevant comments.
  We acknowledge ERC Starting Grant TopoCold for financial support.

\end{document}